# optica

# Near-field coupling of a levitated nanoparticle to a photonic crystal cavity

Lorenzo Magrini,[1] Richard A. Norte,[2] Ralf Riedinger,[1] Igor Marinković,[2] David Grass,[1] Uroš Delić,[1] Simon Gröblacher,[2,3] Sungkun Hong,[1,4] AND Markus Aspelmeyer[1,5]

[1]Vienna Center for Quantum Science and Technology (VCQ), Faculty of Physics, University of Vienna, A-1090 Vienna, Austria
[2]Kavli Institute of Nanoscience, Department of Quantum Nanoscience, Delft University of Technology, 2628CJ Delft, The Netherlands
[3]e-mail: s.groeblacher@tudelft.nl
[4]e-mail: sungkun.hong@univie.ac.at
[5]e-mail: markus.aspelmeyer@univie.ac.at

Quantum control of levitated dielectric particles is an emerging subject in quantum optomechanics. A major challenge is to efficiently measure and manipulate the particle's motion at the Heisenberg uncertainty limit. Here we present a nanophotonic interface suited to address this problem. By optically trapping a 150 nm silica particle and placing it in the near field of a photonic crystal cavity, we achieve tunable single-photon optomechanical coupling of up to $g_0/2\pi = 9$ kHz, three orders of magnitude larger than previously reported for levitated cavity optomechanical systems. Efficient collection and guiding of light through the nanophotonic structure results in a per-photon displacement sensitivity that is increased by two orders of magnitude compared to conventional far-field detection. The demonstrated performance shows a promising route for room temperature quantum optomechanics.

https://doi.org/10.1364/OPTICA.5.001597

## 1. INTRODUCTION

Optical tweezers provide a remarkably simple, yet versatile platform for studying a plethora of intriguing problems in single molecule biophysics [1,2], thermodynamics [3–6], sensing [7,8], or fundamental physics [9,10]. Realizing full quantum control of trapped nanoparticles will enable new insights into quantum-enhanced precision metrology as well as into fundamental aspects of quantum physics [11,12]. The past few years have witnessed rapid progress towards the quantum regime of optically levitated nanoparticles through cavity- [13–16] and feedback-assisted control schemes [17–20]. The primary limitations lie either in small optomechanical coupling strengths to the cavity field, or, for the case of optical tweezers, in significant losses in the detection channel. As every scattered photon induces backaction noise on the particle motion, it is crucial not to "lose" any information carried by light [21], especially in the regime where photon recoil is the dominant source of decoherence. Nanophotonic structures can provide a solution to these problems. Their small mode volumes and high quality factors result in strong optomechanical coupling [22,23]. These nanostructures can also be easily interfaced with a single-mode fiber, hence allowing for efficient collection and guiding of the light from the cavity [24]. Previously, optical nano-devices have been used, for example, to show strong coupling and super-radiance of trapped atoms [25,26], emission rate control of solid state quantum emitters [27,28], label-free single molecule detection [29], or trapping of colloidal particles in liquid [30].

Here we use a nanophotonic cavity to efficiently couple the 3D mechanical motion of a levitated nanoparticle to a single optical mode. Specifically, by placing the particle at a distance of ∼310 nm from a photonic crystal cavity, and exploiting the dispersive coupling to the evanescent component of the strongly confined cavity field, information about the mechanical displacement is encoded into phase fluctuations of the cavity mode [22]. This signal is efficiently outcoupled and guided through single-mode fibers to the detector, resulting in a real-time measurement of the particle motion at high bandwidth and high sensitivity. Our approach therefore complements previous experiments involving nanophotonic structures and colloidal particles, in which the structures are used mainly to trap the particle or to detect the presence of the particle without monitoring its precise position or motion [29,30].

## 2. METHODS

Our experimental setup consists of an optical tweezer and a silicon nitride (SiN) photonic crystal cavity [Fig. 1(a)], both of which are situated inside a vacuum chamber. The cavity is impedance matched, with a fundamental resonance wavelength of $\lambda_{cav} = 1538.72$ nm and an optical loss rate of $\kappa/2\pi = 5.0$ GHz. The input/output mirror is adiabatically transitioned into a tapered waveguide that is interfaced with an open-end tapered fiber [24], yielding a fiber-to-cavity coupling efficiency of $\eta_{\mathrm{cav}} = 0.32$. Taking into account all other losses in the setup, the total detection efficiency of photons approaching the cavity is $\eta = 0.09$ (see Supplement 1). The fiber physically supports the nanocavity by van der Waals forces and can be positioned relative to the optical tweezer using a piezo-actuated three-axis translational stage. The optical tweezer is formed by tightly focusing the laser beam (wavelength $\lambda_{\mathrm{trap}} = 1064$ nm; trap power 150 mW) with a commercial dry objective lens (numerical aperture NA = 0.95) inside the vacuum chamber. The location of the trap within the focal plane is controlled by steering the angle of incidence of the laser at the rear lens of the objective.

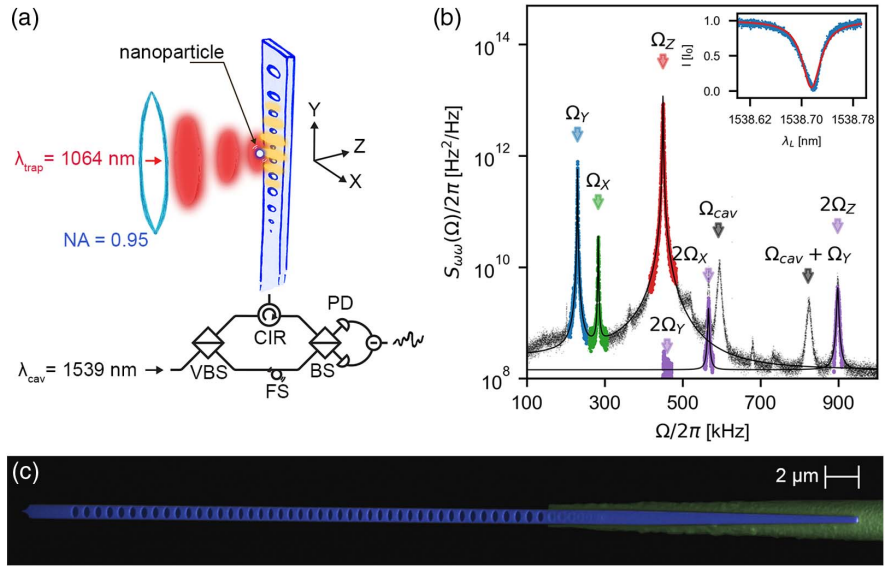

**Fig. 1.** Nanophotonic interface. (a) Sketch of the setup: a dielectric nanoparticle is trapped inside the high intensity lobe formed by the reflection of the optical tweezer light ($\lambda_{\mathrm{trap}} = 1064$ nm) from the surface of the nanophotonic cavity, at a distance of about 310 nm. A laser light resonant with the cavity ($\lambda_{\mathrm{cav}} = 1538.72$ nm) is sent into a variable beam splitter (VBS), which splits it into a weak (260 nW) beam pumping the cavity and a strong (1 mW) local oscillator. The cavity output is redirected by a circulator (CIR) towards a symmetric beam splitter (BS), at which it interferes with the local oscillator. The light in the two output ports is measured using a balanced photo-detector (PD). While the low frequency component of the signal is used to stabilize the interferometer via a fiber stretcher (FS), the high frequency part is directed to a signal analyzer. (b) The measured frequency power spectral density exhibits three mechanical peaks at $\Omega_y/2\pi = 228.3$ kHz (blue), $\Omega_x/2\pi = 280.3$ kHz (green), and $\Omega_z/2\pi = 444.9$ kHz (red). The significantly higher frequency along $z$, which is the direction of the tweezer beam propagation, is caused by the standing wave confinement, and for the radial directions $x$ and $y$, the degeneracy is broken due to the use of polarized light together with tight focusing. Nonlinearities in the trap potential as well as in the optomechanical couplings result in peaks at twice the mechanical frequencies (highlighted in purple). The mechanical vibration of the cavity/fiber assembly at around the frequency $\Omega_{\mathrm{cav}}/2\pi \sim 600$ kHz also induces additional peaks in the spectrum. The inset shows the cavity resonance measured by monitoring the light reflection from the cavity while scanning the pump laser wavelength. The slight asymmetry of the response arises form thermo-optic effects, as we are pumping the cavity at the limit of thermal stability (see Supplement 1). (c) False-colored scanning electron microscope image of the photonic crystal cavity (blue) attached to the tapered fiber (green).

Ultimately, the particle is trapped in a standing wave potential formed by the interference of the focused trapping light with its reflection off the surface of the photonic crystal. To achieve this, we first trap a neutral silica nanoparticle (nominal radius $r = 71.5 \pm 2.0$ nm) with the optical tweezer at ambient pressure [31]. After reducing the pressure to 1.5 mbar, we bring the nanocavity in close proximity to the particle. During this process, the optical trap potential is transformed adiabatically from the single, nominally Gaussian, potential given by the focal spot of the tweezer to the periodic potential induced by the standing wave [25] [Fig. 1(a)]. The locations and actual shapes of the multiple lattice sites are determined by the wavelength of the trap beam and the thickness of the cavity (see Supplement 1 and [25,32]).

Our experimental parameters yield the first minimum of the trapping potential at $z_0 \sim 380$ nm from the device surface, i.e., a surface-to-surface separation between nanosphere and photonic crystal cavity of $d = z_0 - r \sim 310$ nm. Due to the subwavelength transverse dimensions of the nanophotonic device, the cavity field exhibits a considerable evanescent component that decays exponentially with distance. In this region, the displacement of the particle results in a shift of the cavity resonance by $\delta\omega_{\mathrm{cav}} = G_\xi \delta\xi$, where $\xi = x, y, z$ is the direction of mechanical motion and $G_\xi = \partial_\xi \omega_{\mathrm{cav}} \propto \partial_\xi E^* E$ the optomechanical coupling ($E$: evanescent field amplitude). As $G_\xi$ is proportional to the intensity gradient of the cavity field along the direction of motion, each mechanical mode couples to the cavity field with different strength. In particular, the small mode volume results in a large field variation and hence a significantly enlarged coupling when compared to standard levitated optomechanics configurations based on bulk optics [13–15].

## 3. RESULTS AND DISCUSSION

When pumping the cavity on resonance, the position-dependent frequency fluctuation is mapped onto the phase quadrature of the output field, which can then be measured via a shot-noise limited homodyne detection [Fig. 1(a)]. We use this cavity-enhanced measurement to monitor the thermal motion of the trapped particle: the mechanical oscillations in the three spatial directions are observed as distinct frequency components in the homodyne signal [Fig. 1(b)]. Using thermal noise of the particle motion and photon shot noise of the cavity light, we calibrate both displacement and optomechanical coupling (see Supplement 1). We note that, by only injecting 260 nW of optical power into the cavity and at an overall detection efficiency of 9%, we achieve a displacement sensitivity of $(3.3 \pm 0.5) \times 10^{-12}$ m/$\sqrt{\text{Hz}}$, similar to what is measured in far-field detection with 1 mW of detected light. This amounts to an increase in the position sensitivity per photon by more than a factor of 100.

At the optimal position we measure coupling rates along the $z$ direction of motion, i.e., orthogonal to the cavity surface, of $G_z/2\pi = 3.6 \pm 0.4$ MHz/nm. This is consistent with our finite element method (FEM) simulation (see Supplement 1) and corresponds to a single-photon optomechanical coupling $g_0/2\pi \equiv z_{\text{zpf}} G_z/2\pi$ of $9.3 \pm 0.9$ kHz ($z_{\text{zpf}} = (\hbar/2m\Omega_z)^{1/2}$: mechanical zero point fluctuation of the particle motion in the $z$ direction). Another intriguing feature of photonic crystal cavities is the strong spatial variation of the cavity field $E$, which results in a significant position-dependent optomechanical coupling for all three spatial directions of motion.

By changing the particle position relative to the cavity, we can therefore tune the optomechanical coupling of all mechanical modes [33]. We experimentally demonstrate this by scanning the particle position in a plane perpendicular to the $z$ axis while simultaneously monitoring the cavity signal. The observed strong modulations in all three coupling rates are in good agreement with FEM simulations (Fig. 2). As the motion of the levitated nanoparticle represents a subwavelength probe, this measurement allows us to image the 3D intensity gradient of the nanophotonic cavity mode in super-resolution, i.e., not limited by diffraction (Fig. 2). Compared to standard near-field scanning techniques, such as scanning near-field microscopy [34], our resolution is defined by the extent of the particle motion and not by the physical size of the probe. As a consequence, the imaging is fundamentally limited only by the ground state size of the trapped particle, i.e., to a resolution of some picometers. In spite of this, position drifts and the accuracy of our positioner currently limit the imaging resolution to some tens of nanometers [Fig. 2(b)] in a field of view of half a micrometer square.

Our system also enables tunability of the mechanical frequencies without affecting the coupling strength to the cavity field. In other words, we can modify the trapping potential independent of the trapping distance. To demonstrate this, we move the cavity along the $z$ direction, away from the focus of the trapping beam [Fig. 3(a)]. The optomechanical coupling stays constant [Fig. 3(c)], indicating that the relative distance between the particle and the cavity remains unchanged. This behavior can easily be understood when considering the formation of the standing wave by the cavity reflection. The positions of the antinodes are solely determined by the location of the cavity and its thickness, locking the trap position to the cavity. At the same time, the mechanical frequency is reduced because the high divergence of the tightly focused optical tweezer leads to a sharp decrease of the intensity at the trap location [Figs. 3(b) and 3(d)].

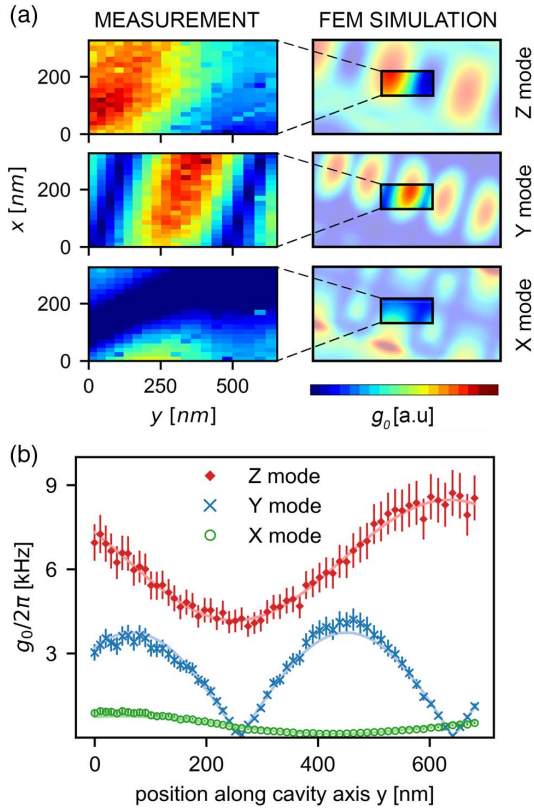

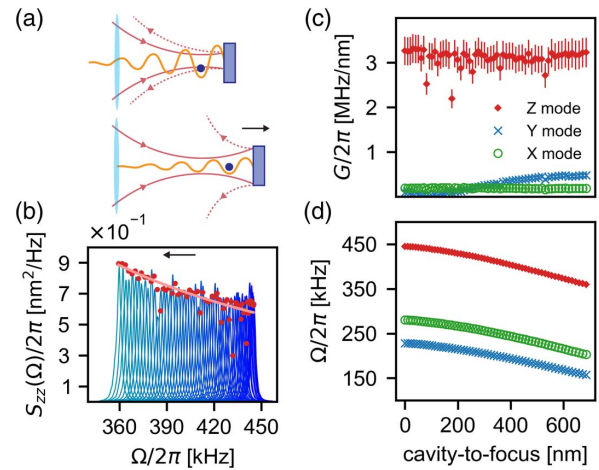

**Fig. 2.** Optomechanical coupling. (a) Measured (left) and simulated (right) intensity map of the single-photon optomechanical coupling rates $g_0$ for the three spatial modes. Because of heating from the tweezer light (see Supplement 1), at every position the cavity is reset on resonance before recording the interferometric signal. (b) Position scan of the single-photon optomechanical coupling rates along the $y$ direction and close to the cavity center for the modes along $x$ (green circles), $y$ (blue crosses), and $z$ (red diamonds). Solid lines are fits based on our cavity field model (see Supplement 1). As the scan was performed slightly off the cavity center, the coupling to the $z$ mode is non-vanishing while we can suppress the $x$ and $y$ couplings. The main contribution to the error bars is given by the uncertainty in the shot-noise level determined by the integration time of $\sim 3$ s.

**Fig. 3.** Position locking. (a) Sketch of the nanoparticle (blue dot), trapped in the standing wave potential (orange) formed by the reflection of the focused tweezer light (red) by the photonic crystal cavity (blue rectangle). The data is taken by moving the photonic crystal along the direction of propagation of the tweezer beam ($z$). While the particle's distance to the cavity remains locked, the divergence of the tweezer causes a reduction of the trapping potential. (b) Position power spectral density for the $z$ mode $S_{zz}(\Omega)$ (blue) measured as cavity-focus increases (in direction of the arrow). The variance of the motion given by the peak integral (red dots $\propto \int S_{zz}(\Omega)\mathrm{d}\Omega$) changes with the mechanical frequency as stated by the equipartition theorem (pink solid line $\propto 1/\Omega_z^2$). Deviation from the expected Lorentzian peak is given by the fluctuations during the integration time, which effectively reduce the peak height. (c) Frequency shift per displacement $G$ plotted as a function of the cavity distance to the focal plane, for the $z$ mode (red diamonds), $y$ mode (blue crosses), and $x$ mode (green circles). (d) Mechanical frequencies for the three modes as a function of the cavity distance to the focal plane.

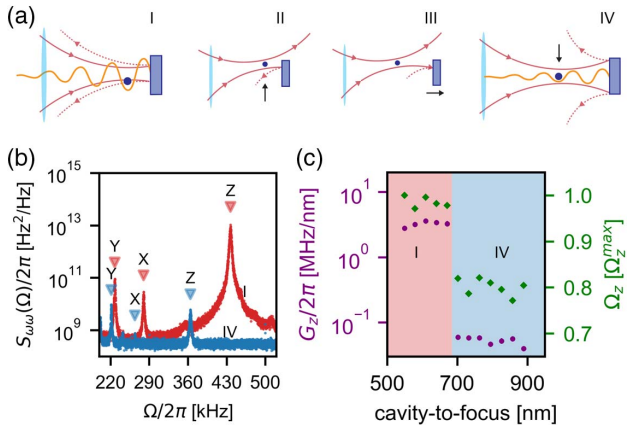

**Fig. 4.** Loading of the particle into the lattice. (a) The particle is initially trapped in the closest of the cavity trap sites (I). We steer the tweezer away from the cavity (II) and subsequently change the cavity position (III). Finally the particle is steered back in front of the cavity (IV). Depending on the cavity-to-focus distance, the particle will slide into different sites. (b) Frequency power spectral densities measured in the case of the particle being in the first trapping site (red, I) or in the second (blue, IV). The small unlabeled peak in the blue spectrum is an electronic noise peak common to all measurements. (c) Optomechanical coupling (purple dots) and mechanical frequency (green diamonds) for the z mode as a function of the initial cavity-to-focus distance.

Finally, we demonstrate reliable, deterministic loading of the nanoparticle into the different standing wave optical lattice sites. This is achieved by a sequence of optical tweezer and cavity position control steps [Fig. 4(a)]. We first terminate the standing wave by moving the particle to the side of the photonic crystal cavity. After displacing the cavity along the z axis, the particle is moved back and the standing wave is reestablished. When the cavity is sufficiently displaced, the particle slides into the next trap location of the reappearing standing wave. We observe this behavior when the cavity displacement is greater than $\lambda_{trap}/4$ [Fig. 4(c)]. At this second trap location, the optomechanical coupling rate is reduced by two orders of magnitude, consistent with FEM simulation [see Fig. 4(b) and Supplement 1].

## 4. CONCLUSIONS

In summary, we have realized a low-loss and widely tunable hybrid optomechanical system combining optical levitation of a nanoparticle with a nanophotonic cavity via near-field coupling. The displacement sensitivity per photon of our platform is more than two orders of magnitude higher than what was shown using far-field detection [19]. This opens a direct route for quantum feedback control. Specifically, ground state cooling with feedback requires $\eta > (1 + 1/C_q)/9$ with $C_q$ the quantum cooperativity [35–37], yielding a minimally required value for the detection efficiency of $\eta > 1/9 \approx 0.11$. While far-field detection is currently limited at $\eta \sim 10^{-3}$ [19], we here demonstrate $\eta = 0.09$, i.e., already close to the required bound. We anticipate that a more stringent screening process over multiple cavity transfer trials (see Supplement 1) will yield fiber-cavity assemblies with coupling efficiency exceeding $\eta_{cav} = 0.96$, as was previously shown by Burek et al. [24]. It would result in an overall detection efficiency of $\eta > 0.3$.

The other relevant figure of merit for quantum state control is the quantum cooperativity $C_q = (4g_0^2 n_{cav})/(\kappa \Gamma_m n_{th})$, where $n_{cav}$ ($n_{th}$) and $\kappa$ ($\Gamma_m$) are the cavity photon (mechanical phonon) occupation and loss rate, respectively [38]. Our current value ($C_q \sim 10^{-9}$) is mainly limited by the fact that, in the absence of feedback stabilization of the particle, the operating pressure cannot be decreased below ∼1 mBar (corresponding to mechanical loss rates $\Gamma_m/2\pi$ of more than $10^3$ Hz). Implementing stable feedback cooling will allow us to reach ultra-high vacuum levels ($10^{-8}$ mbar and below) at which mechanical losses are limited by photon recoil to $\Gamma_m/2\pi \approx 10^{-4}$ Hz. This will result in an immediate improvement of cooperativity by more than seven orders of magnitude. In the present configuration, the main bottleneck is the mechanical support of the cavity, which causes alignment drifts and hence limits feedback particle stabilization in ultra-high vacuum. One workaround will be to use rigidly mounted on-chip (instead of fiber supported) cavities. This will also improve the thermal anchoring of the cavity and therefore enable a higher intra-cavity photon number $n_{cav}$, which is now limited to $n_{cav} \sim 800$ because of thermo-optic heating. With a more careful design and fabrication, the cavity optical losses $\kappa/2\pi$ can be reduced to as low as 20 MHz in silicon [39] and 1 GHz in SiN [40]. The cavity thickness directly affects the boundary condition for the standing wave trap formation such that, with an appropriately chosen thickness, the particle can be trapped within 200 nm from the cavity surface (see Supplement 1 and [25]). This would result in an increase of the optomechanical coupling rate by one order of magnitude. Incorporating all these improvements will allow us to achieve $C_q > 10$ and thus place the system deep into the strong cooperativity regime. This will enable a new generation of chip-based levitated quantum sensors operating at room temperature. For example, the high bandwidth of our system ($\kappa \gg \Omega_m$) makes it an ideal platform for implementing measurement-based quantum state preparation using pulsed interactions [41], which is a complementary approach to quantum control methods based on cavity sideband driving [38]. The high coupling and relatively low frequencies place the system in reach of the nonlinear optomechanical regime ($g_0 \approx \Omega_m$) [42]. Exploiting the design capabilities for the spatial modes in photonic crystal cavities, our system can also be used for studying effects of self-induced backaction and non-harmonic dynamics in both the classical and quantum regimes [43]. Also, the expected force noise of $10^{-20}$ N/$\sqrt{Hz}$ will allow a detailed study of short-range surface forces at sub-micrometer distances [7–9].

**Funding.** Austrian Science Fund (FWF) (F40, P28172, W1210); Stichting voor Fundamenteel Onderzoek der Materie (FOM) (15PR3210); Nederlandse Organisatie voor Wetenschappelijk Onderzoek (NWO), as part of the Frontiers of Nanoscience program, as well as through a Vidi grant (680-47-541/994); H2020 European Research Council (ERC) (QLev4G 649008, Strong-Q 676842); H2020 Marie Curie ITN (OMT 722923); H2020 QuantERA ERA-NET Cofund

in Quantum Technologies (TheBlinCQ EP/R044082); Vienna Doctoral School of Physics (VDS-P).

**Acknowledgment.** We thank Ramon Moghadas Nia for valuable lab support; Eugene Straver, Michael J. Burek, and Marko Lončar for their technical advice on the tapered fibers; and Nikolai Kiesel and Lukas Novotny for helpful discussions. L. M. is supported by the Vienna Doctoral School of Physics (VDS-P); R. R. is a recipient of a DOC fellowship of the Austrian Academy of Sciences at the University of Vienna; and L. M., R. R., D. G., and U. D. are supported by the FWF under project W1210 (CoQuS).

See Supplement 1 for supporting content.